\documentclass[twocolumn,preprintnumbers,amsmath,amssymb,nofootinbib]{revtex4}
 \usepackage{graphicx}

\usepackage{graphicx}
\usepackage{dcolumn}
\usepackage{enumerate}
\newcolumntype{=}{D{=}{=}{10.10} }

\begin{document}

\title{A compact design for a magnetic synchrotron to store beams of hydrogen atoms}

\author{Aernout P P van der Poel$^1$}
\author{Katrin Dulitz$^2$\footnote{Current address: Laboratory of Physical Chemistry, ETH Z{\"u}rich, Vladimir-Prelog-Weg 2, 8093 Z{\"u}rich, Switzerland}}
\author{Timothy P Softley$^2$}
\author{Hendrick L Bethlem$^1$}
\email{H.L.Bethlem@vu.nl}
\affiliation{$^1$ LaserLaB, Department of Physics and Astronomy, VU University Amsterdam, De Boelelaan 1081, 1081 HV Amsterdam, The Netherlands}
\affiliation{$^2$ Department of Chemistry, University of Oxford, Chemistry Research Laboratory, 12 Mansfield Road, Oxford, OX1 3TA, United Kingdom}

\begin{abstract}
We present a design for an atomic synchrotron consisting of $40$ hybrid magnetic hexapole lenses arranged in a circle. We show that for realistic parameters, hydrogen atoms with a velocity up to $600$\,m/s can be stored in a $1$-meter diameter ring, which implies that the atoms can be injected in the ring directly from a pulsed supersonic beam source. This ring can be used to study collisions between stored hydrogen atoms and molecular beams of many different atoms and molecules. The advantage of using a synchrotron is two-fold: (i) the collision partners move in the same direction as the stored atoms, resulting in a small relative velocity and thus a low collision energy, and (ii) by storing atoms for many round-trips, the sensitivity to collisions is enhanced by a factor of 100-1000. In the proposed ring, the cross-sections for collisions between hydrogen, the most abundant atom in the universe, with any atom or molecule that can be put in a beam, including He, H$_2$, CO, ammonia and OH can be measured at energies below $100$\,K. We discuss the possibility to use optical transitions to load hydrogen atoms into the ring without influencing the atoms that are already stored. In this way it will be possible to reach high densities of stored hydrogen atoms.
\end{abstract}

\maketitle

\section{Introduction}
\label{sec:introduction}

The ability to control the translational energy and the energy spread of a molecular beam enables collision studies that probe molecular interaction potentials in great detail~\cite{vandeMeerakker2012,Narevicius2012,Brouard2014}. A holy grail of this research is to be able to observe resonances in the collision cross-section as a function of collision energy. These resonances occur when the kinetic energy of two colliding molecules is converted into rotational energy as a result of the anisotropy of the potential energy surface (PES). This leaves the molecules with insufficient translational energy to overcome the Van der Waals attraction, transiently binding the molecules together~\cite{Balakrishnan2000}. These long-lived excitations of the collision complex appear as sharp resonances in graphs of the cross-section as a function of the collision energy. The position and width of these resonances provide an extremely sensitive probe of the interaction potential energy surfaces~\cite{Zare2006,Chandler2010}. Since these resonances arise from the rearrangement of rotational energy, they do not occur for cold atom-atom collisions. Another motivation for research into cold collisions is the fact that, at low temperatures, the collision process becomes sensitive to externally applied electric or magnetic fields, which gives a handle to control and steer the outcome of a chemical reaction~\cite{Krems2005,Krems2008}.

The crossed-molecular-beam method -- pioneered by Dudley Herschbach and Yuan T. Lee -- has yielded a detailed understanding of how molecules interact and react~\cite{Herschbach1987,Lee1987}. In this method, beams of atoms and molecules in high vacuum are crossed at right angles and made to collide. Inhomogeneous electric and magnetic fields can be used to prepare the molecules in selected quantum states and to orient them before the collision, while laser techniques in combination with sophisticated ion optics can be used to determine both the quantum state of the products and their kinetic energy. The collision energy in a crossed-beam experiment is determined by the velocity of the molecular beams and is typically around $1000$\,cm$^{-1}$. Therefore, while there is plenty of data available at high energy, experimental data for low-energy collisions is scarce.

Recently, collision studies have been performed between Stark-decelerated beams of OH radicals and conventional beams of rare gas atoms in a crossed-beam geometry. In these experiments, the total collision energy is tuned from $100$ to $600$\,cm$^{-1}$ by varying the velocity of the OH radicals, while the velocity of the rare-gas beam is kept fixed. In this way the threshold behavior of state-to-state inelastic scattering cross-sections was accurately determined~\cite{Gilijamse2006,Scharfenberg2010,Scharfenberg2011}. Experiments have also been conducted using collisions between Stark-decelerated OH molecules and a state-selected beam of NO molecules~\cite{Kirste2012}. 

Low collision energy can also be attained by performing crossed-beam experiments at low crossing angle~\cite{Toennies1979}. In recent experiments, scattering resonances were observed in collisions between O$_{2}$ or CO and hydrogen molecules at energies between $5$ and $30$\,K~\cite{Chefdeville2012,Chefdeville2013}.  Even lower temperatures can be obtained by using magnetic~\cite{Henson2012,Lavert-Ofir2014} or electric guides~\cite{Jankunas2014a,Jankunas2014} to merge two molecular beams into a single beam. The collision angle is zero in this case, so that the relative velocity of the reactants in the centre-of-mass frame becomes simply the difference between the two beam speeds. In contrast to a crossed-beam experiment this difference can become zero even at high beam velocities, thus rendering the slowing of the molecules unnecessary. Until now, experiments with merged supersonic beams have been restricted to penning ionization collisions of metastable He or Ne atoms.

In a different approach, trapped ions at millikelvin temperatures are monitored while a slow beam of molecules passes through the trap to study reactive ion-molecule collisions~\cite{Willitsch2008a,Chang2013}. The advantage of this method is that the ions are stored for a long time while their number can be accurately determined, allowing the detection of reactions even when they occur at a rate of one per minute. In a similar fashion, collisions of slow beams of ammonia with magnetically trapped OH molecules were observed~\cite{Sawyer2011}.  

Here, we discuss the possibility to study collisions between hydrogen atoms stored in a magnetic synchrotron and supersonic beams of different atoms and molecules, combining the virtues of the two approaches mentioned above:
\begin{enumerate}[(i)]
\item	The collision partners move in the same direction as the stored molecules, resulting in a small relative velocity and thus a low collision energy. 
\item The sensitivity to collisions is enhanced by orders of magnitude by storing the atoms during many round-trips.
\end{enumerate}
We present a design for a synchrotron consisting of $40$ hybrid magnetic hexapole lenses arranged in a circle. We show that, for realistic parameters, a significant fraction of a supersonic beam of hydrogen atoms with a velocity of $600$\,m/s can be directly loaded into a $1$-meter diameter ring.


\section{Storage rings for neutral particles}
\label{sec:storagerings}

In its simplest form, a storage ring is a trap in which the particles possess a minimum potential energy on a circle rather than having a minimum potential energy at a single location in space. The advantage of a storage ring over a trap is that packets of particles with a non-zero mean velocity can be confined. While circling the ring, these particles can be made to interact with other particles repeatedly, at well-defined times and at distinct positions~\cite{Crompvoets2001}. A simple storage ring for polar molecules or paramagnetic atoms and molecules can be devised by bending a magnetostatic or electrostatic hexapole guide into a torus, such as the $1$m-diameter magnetic storage ring used by K{\"u}gler~\emph{et al.} to store cold neutrons \cite{Kugler1985} or the $0.25$m-diameter electrostatic storage ring used by Crompvoets \emph{et al.} to store Stark-decelerated ammonia molecules \cite{Crompvoets2001}. A strong-focusing storage ring for slow (100\,m/s) atomic hydrogen in high-field seeking states was proposed by Thompson~\emph{et al.}~\cite{Thompson1989}.  

In a storage ring, an injected packet of atoms and molecules will gradually spread out along the ring as a result of its longitudinal velocity spread, eventually filling the entire ring. To fully exploit the possibilities offered by a ring structure, it is necessary that the particles remain in a bunch as they revolve around the ring, ensuring a high density of stored particles and making it possible to inject multiple packets into the ring without affecting the packet(s) already stored. Thus, in addition to the transverse forces that focus the particles and keep them on a circular orbit, we need to apply a force in the longitudinal direction. In modern-day particle accelerators, the different tasks are performed by separate elements (RF-cavities for bunching, magnetic quadrupoles for transverse focusing, and dipole magnets for bending). An elaborate design of a synchrotron for polar molecules in low-field and high-field seekings states based on this approach was presented by Nishimura~\emph{et al.} \cite{Nishimura2003}. Because of its low degree of symmetry, however, this design is seriously hindered by non-linearities in the forces experienced by the molecules. 

A different approach was taken by Heiner~\emph{et al.} who demonstrated a synchrotron consisting of two hexapoles which were bent into semi-circles and separated by a short gap~\cite{Heiner2007}. By appropriately switching the voltages applied to the electrodes as the molecules pass through the gaps between the two half rings, molecules experience a force that keeps them together in a compact bunch. In this design, the necessary bunching, bending and focusing forces are thus delivered by a single element. This allows for a much more compact design of high symmetry that is much less sensitive to non-linearities. A synchrotron consisting of $40$ straight electric hexapoles arranged in a circle with a radius of $250$\,mm was subsequently demonstrated by Zieger~\emph{et al.}~\cite{Zieger2010a}. In this ring, packets of bunched molecules were observed to make up to $1025$ round-trips, thereby passing a distance of over 1\,mile. Up to 26 molecular packets  were stored simultaneously in this ring structure; $13$ revolving clockwise and $13$ revolving counter clockwise~\cite{Zieger2013a}.

In this paper, we discuss the feasibility of a magnetic synchrotron. Compared to electric fields, magnetic fields offer many technical challenges but also some new possibilities. A synchrotron consisting of 40 electromagnetic hexapoles, the exact counterpart of the electric synchrotron discussed above, would demand an impractically large cooling capacity owing to Ohmic heating caused by the required application of very high currents. However, large magnetic fields can also be produced by permanent magnets. Nowadays, inexpensive magnets made from rare-earth metals with a surface magnetization of up to $1.4$\,T are commercially available in various sizes. Such magnets have been used in a number of experiments to manipulate paramagnetic atoms and molecules. For instance, curved quadrupole, hexapole and octupole guides based on permanent magnets in a Halbach configuration~\cite{Halbach1980} have been used to manipulate slow atomic beams ~\cite{Ghaffari1999,Goepfert1999,Nikitin2003,Beardmore2009}. By extending such a guide, a simple magnetic storage ring can be devised. However, methods are needed (i) to allow loading atoms into such a ring and (ii) for bunching the atoms stored in it. Here, we present a design for a synchrotron that uses hybrid hexapole lenses to combine the strong magnetic fields offered by permanent magnets with the flexibility of electromagnets. 

The paper is organized as follows. In section~\ref{sec:lens}, we will describe the magnetic field resulting from a single hybrid lens and show that, by changing the currents through the electromagnets, we can provide all necessary fields for loading, bending, focusing and bunching. In section~\ref{sec:synchrotron}, we calculate the transverse and longitudinal acceptance for a ring composed of $40$ of these lenses arranged in a circle with a diameter of $1$\,meter. Finally, in section~\ref{sec:collisions} we discuss how we plan to use this synchrotron for studying collisions between atomic hydrogen and various atoms and molecules at low temperatures.


\section{A hybrid magnetic hexapole}
\label{sec:lens}

\begin{figure*}[t!]
\begin{center}
\includegraphics[width=\linewidth]{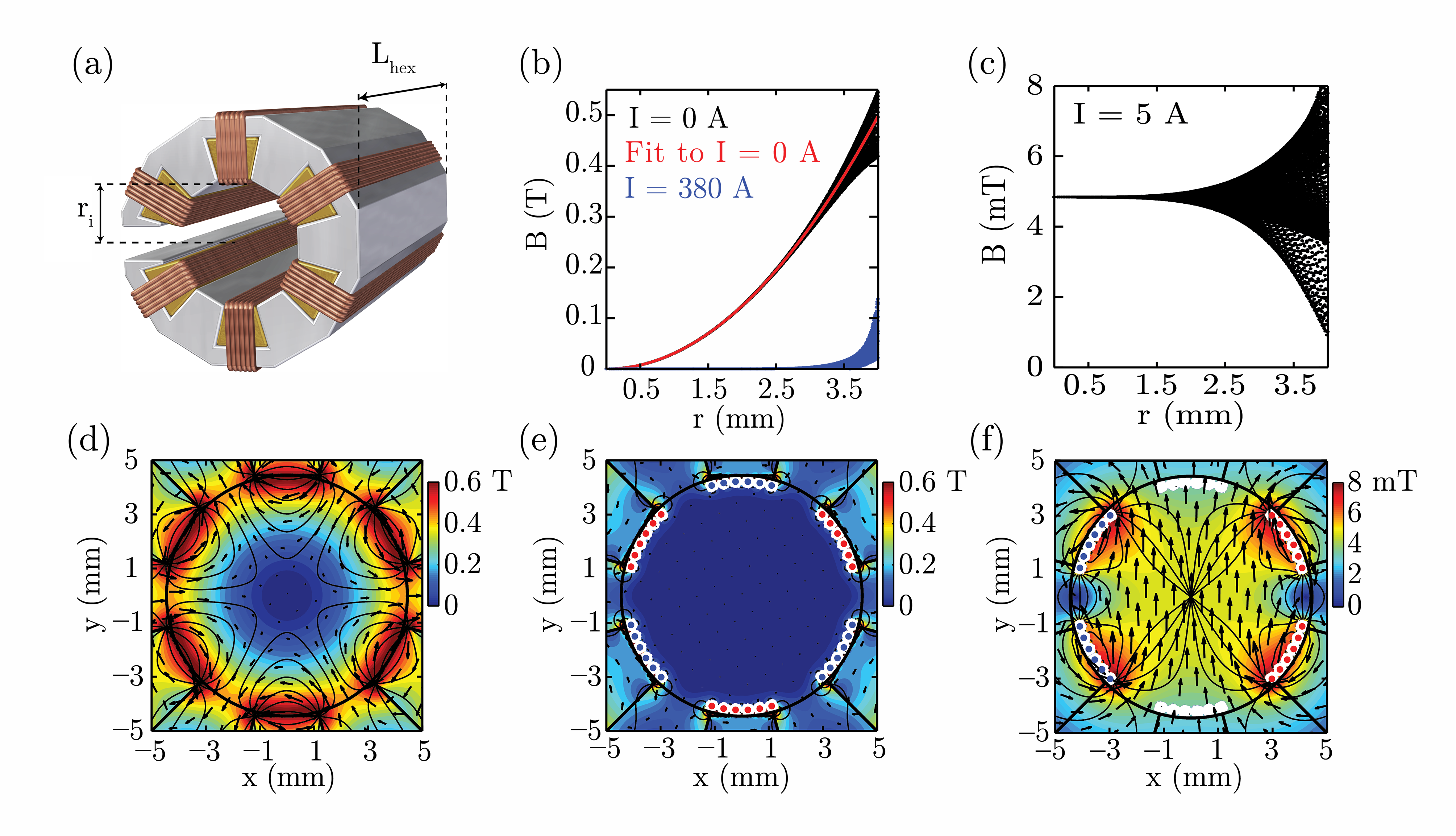}
\end{center}
\caption{(a) Schematic drawing of a hybrid lens composed of six permanent magnets and six sets of six wires held by an aluminium housing. The inner radius of the lens is $r_{i}$ = $4$\,mm and a $1.8$mm-wide slit allows atoms to be injected off axis. 
(b) Transverse magnetic field in the centre of the hexapole assembly (at all angles $\phi$) for the configurations in (d) and (e), in which currents of 0\,A and 380\,A are applied to the electromagnets, respectively. The simulated data for the permanent-magnet arrangement is well-approximated by a quadratic dependency to $a_{H}r^2$ (red line, yielding $a_{H}$~=~0.03~T/mm$^2$), where $a_{H}$ is the magnetic field curvature.
(c) Transverse magnetic field in the centre of the hexapole assembly (at all angles $\phi$) for configuration (f). 
(d) Density plot of the magnetic field magnitude $B$ (in T) and flux lines (using $B_{\phi}$) in the centre of an assembly of six permanent magnets in hexapole configuration, obtained from simulations in Radia 4.29.
(e) Same as in (d) but superimposed with a magnetic hexapole field from the six sets of current-carrying wires (filled white circles) operated at a current of $380$\,A, such that the magnetic field of the permanent magnets is nearly canceled out. Red (blue) dots indicate that the current is directed into (out of) the plane of projection.
(f) Density plot of the magnetic field magnitude $B$ (in T) and flux lines (using $B_{\phi}$)  for the wires in a dipole configuration. Filled red (blue) dots and open red (blue) dots indicate that currents of $5$\,A are directed towards (out of) the plane of projection, respectively.}
\label{fig:lens}
\end{figure*}

Figure~\ref{fig:lens}(a) shows a schematic drawing of a hybrid lens composed of six permanent magnets and six sets of six current-carrying wires held by an aluminum housing. The inner radius of the lens is chosen to be $r_{i}$ = $4.45$\,mm. A $1.8$\,mm wide slit allows atoms to be injected off axis.

The permanent magnets are made from NdFeB and have a remanence of $B_0 = 1.3$\,T. The magnetization vector of each magnet is rotated by $180$ degrees with respect to its neighboring magnet to approximate a hexapole field. Figure~\ref{fig:lens}(d) shows a density plot of the magnetic field magnitude $B$ (in T) and flux lines (using $B_{\phi}$) obtained from simulations in Radia 4.29~\cite{Chavanne2009} when the current through the coils is zero. In figure~\ref{fig:lens}(b), the black curves (denoted by $I=0$\,A in the legend) show the radial dependence of this field for different angles $\phi$. The red curve shows a quadratic fit, $B=a_{H}r^{2}$, to the calculated magnetic fields, yielding $a_{H}=0.031$\,T/mm$^{2}$. Note that adding six more magnets, with their magnetization pointing alternately towards or away from the centre, results in a field that is not only a better approximation to a hexapole field but also two times stronger. Unfortunately, such a field cannot be used as the magnets would block the injection beam.

The permanent-magnet array provides the necessary force to keep atoms confined in the ring. In order to inject new packets into the ring, it is necessary to temporarily reduce the field, which is done by applying a current pulse through the six sets of $0.45$\,mm thick wires that are wound around the aluminium housing. Figure 1(e) shows the magnetic field magnitude and flux lines inside the lens when a current of $380$\,A runs in an alternating direction through the six coils. The radial dependence of this field for different angles $\phi$ is shown by the blue curves in figure 1(b) (denoted by $I=380$\,A in the legend). As observed, the magnetic field of the permanent magnets is almost completely canceled by the electromagnet, illustrating that both fields have a very similar shape. Note that the field far away from the hexapole (not shown) is sufficiently small that it would not deflect the injection beam significantly.

In order to apply a focusing force in the longitudinal direction of the ring to bunch the atoms, we superimpose an oscillating dipole field to the confinement field provided by the permanent magnets. Such a field is created by running opposite currents through the two sets of wires to the left and to the right. When a current of $5$\,A is applied, a near uniform field with a magnitude of $B=5$\,mT is generated, as shown in figure 1(f). The corresponding radial dependence is shown by the curves in figure 1(c) (denoted by $I=5$\,A). By oscillating this field synchronously with the velocity of the stored packet, atoms can be accelerated or decelerated without influencing the transverse motion. Note that by adding the dipole field to the confinement field, the magnetic field zero is displaced from the geometric centre of the hexapole. By oscillating the dipole field, the zero-point is oscillating around the geometrical centre. As the atoms need to be in sufficiently high magnetic fields to prevent Majorana transitions, the maximum bunching field that can be applied is limited. At $5$\,A the point of zero magnetic field moves by about $0.4$\,mm. 

As permanent magnets are magnetized by applying a high field to them, one may worry whether the currents used for bunching and injection will demagnetize them. For NdFeB material at room temperature, the applied B field must be quite high to induce a significant decrease of the magnetization~\cite{Arnold}, therefore the effect of the bunching field on the permanent magnets is negligible. However, the field required for injection might lead to a slight demagnetization. If this turns out to be significant, this can be compensated by running a small current in the opposite direction to increase the magnetic field. Alternatively the magnets may be remagnetised by applying a strong current pulse in the opposite direction as used for injection.

\section{A synchrotron consisting of $40$ hybrid magnetic hexapoles}
\label{sec:synchrotron}

In this section, we will estimate the transverse and longitudinal trapping frequencies and the acceptance of a synchrotron consisting of many lenses of the type discussed in the previous section, using models that were developed to analyse the motion through electric storage rings and synchrotrons \cite{Crompvoets2001,Heiner2007,Zieger2010a}. Although these models are based on a number of approximations and simplifications, the high degree of symmetry of the ring ensures that the values obtained are sufficiently reliable.

\subsection{Transverse motion in a hexapole torus}
\label{sec:transverse1}

Let us start by considering a storage ring obtained by bending a linear hexapole guide into a torus with a radius $R_{\mathrm{ring}}$. A hydrogen atom in a low-field seeking state inside this torus will experience two forces; a force directed towards the centre that results from the inhomogeneous magnetic field and a centrifugal force that depends on its forward velocity. These forces cancel at a certain radius which we will refer to as the \textit{equilibrium radius}, $r_{\mathrm{equi}}$:

\begin{equation}
\frac{mv_{\phi}^{2}}{R_{\mathrm{ring}}+ r_{\mathrm{equi}}} = -\mu\frac{dB}{dr}(r_{\mathrm{equi}}) 
= - 2\mu a_{H}r_{\mathrm{equi}},
\label{eq:force}
\end{equation} 

\noindent where $m$ is the mass of the hydrogen atom, $v_{\phi}$ its longitudinal velocity, and $\mu$ the magnetic moment that is in general a function of the magnitude of the magnetic field. If we assume that $r_{\mathrm{equi}} \ll R_{\mathrm{ring}} $, and neglect the hyperfine structure splitting, i.e., if we assume that the Zeeman shift is linear even in zero field, we find

\begin{equation}
r_{\mathrm{equi}} 
\approx  \frac{mv_{\phi}^{2}}{2  R_{\mathrm{ring}} \mu_{B} a_{H}}  
= \frac{v_{\phi}^{2}}{ R_{\mathrm{ring}}  \omega_{r}^2},  
\label{eq:equilibriumorbit}
\end{equation} 

\noindent where $\omega_{r}$ is the betatron (radial) frequency and $\mu_{B}$ is the Bohr magneton, $1.4$\,MHz/Gauss. If an atom with a certain longitudinal velocity starts at the radius that is appropriate for its speed, it will form a closed orbit with a constant radial position while it revolves around the ring. Atoms flying with the same forward velocity but with a different radial position, or with a non-zero radial velocity, will oscillate around this hypothetical atom with the betatron frequency. The amplitude of this motion is limited by the aperture of the hexapole. For higher velocities, the equilibrium orbit will move closer to the magnets, and the maximum amplitude of atoms oscillating around the equilibrium orbit is correspondingly decreased. The acceptance of the ring, the maximum area in phase-space that the atoms may occupy in order to be stably confined, is thus given by 

\begin{equation}
\mathrm{Acceptance} = \pi\omega (r_i - r_{\mathrm{equi}})^2 ,
\label{eq:acceptance}
\end{equation}

\noindent
with $r_{i}$ the inner radius of the hexapole lens. In a ring made by bending the hexapole guide as discussed in section~\ref{sec:lens}, the betatron frequency is $\sim 3$\,kHz. This implies that a hydrogen atom with a longitudinal velocity of $600$\,m/s will have an equilibrium radius of $2$\,mm in a torus of $1$\,meter diameter. The radial acceptance at $600$\,m/s is an ellipse with (full) axes of $4$\,mm and $76$\,m/s.

\subsection{Transverse motion in a segmented ring}
\label{sec:transverse2}

\begin{figure*}[t!]
\begin{center}
\includegraphics[width=\linewidth]{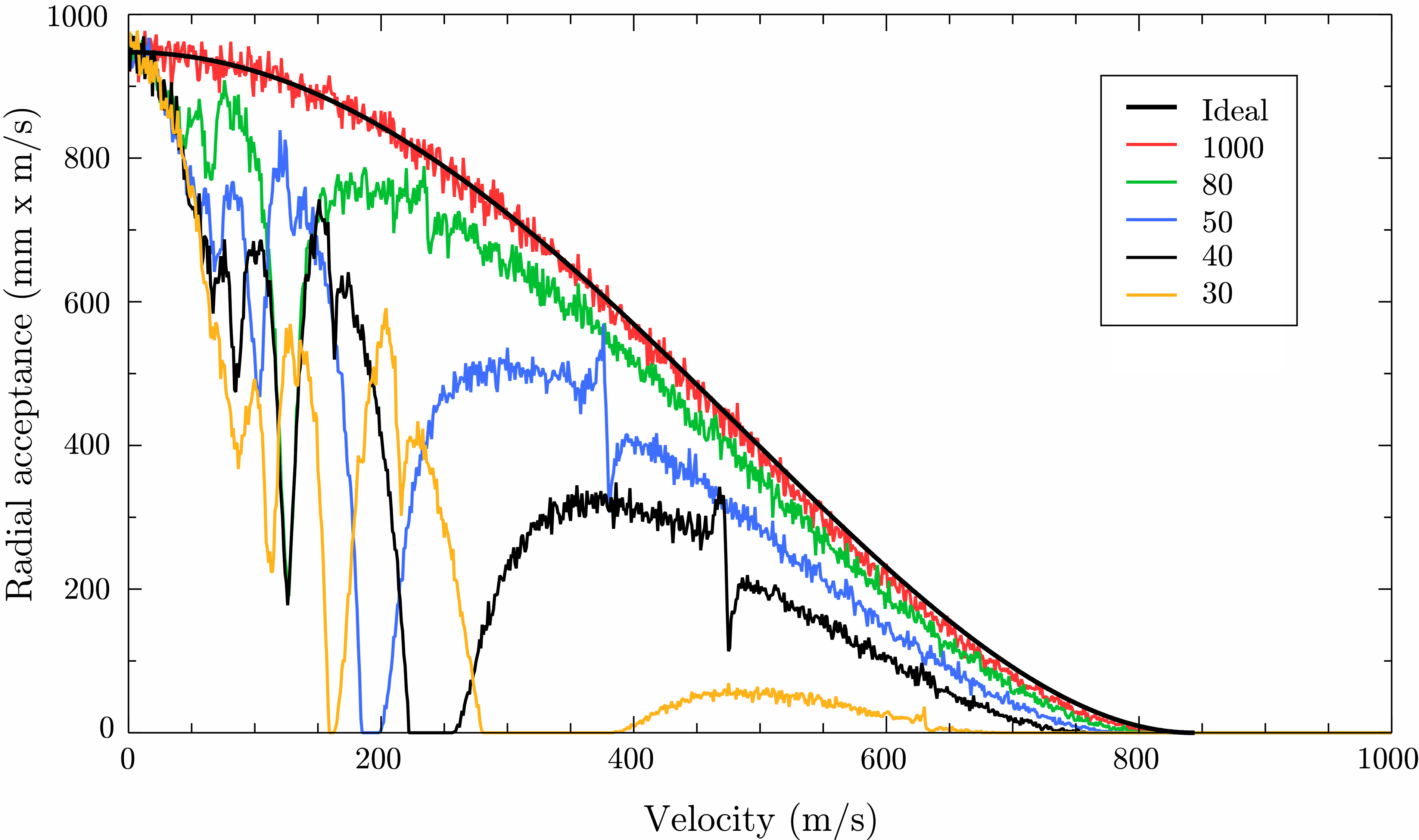}
\end{center}
\caption{The radial acceptance of a ring composed of $30$, $40$, $50$, $80$, and $1000$ segments is plotted as a function of forward velocity. A ring consisting of $1000$ straight hexapoles is essentially identical to the ideal case (bold black line). Reducing the number of segments reduces the acceptance of the ring and additionally introduces stop-bands.}
\label{fig:acceptance}
\end{figure*}

In order to keep the atoms in a tight bunch as they revolve around the ring, and to be able to inject multiple packets, it is necessary to break the symmetry of the ring, i.e., to make a ring composed out of many segments separated by short gaps. Rather than curved hexapoles, we consider the use of straight hexapoles as these are much simpler to construct. In this section, we will use Monte-Carlo simulations to determine how the acceptance of a ring depends on the number of segments. Similar calculations for an electric synchrotron were presented in \cite{Heiner:Thesis,Zieger2013a}.

In our simulation, the magnetic force on the atoms in the hexapole is assumed to be perfectly linear with a force constant of $k_{\mathrm{hex}} = m \omega_r^2$, where $m$ is the mass of the atom and $\omega_r/2\pi$ is the betatron frequency, taken to be $3$\,kHz. Using this force as an input, the trajectories of the atoms are calculated using a second-order Runge-Kutta method. At every time step, it is checked whether or not the distance of the atom with respect to the centreline of the hexapole in which it is located (at that instant), exceeds the aperture $r_i$ of the hexapole; $r_i$ = 4\,mm in our simulations. The two-dimensional acceptance for a given longitudinal velocity and number of ring-segments is found from the fraction of molecules that remains in the ring after $20$\,ms, multiplied by the initial phase-space area.

Figure~\ref{fig:acceptance} shows the resulting acceptance for a ring composed of $30$, $40$, $50$, $80$, and $1000$ straight hexapoles, whose lengths are adjusted accordingly to accommodate the $1$meter diameter of the ring. For each data point, $4000$ molecules are generated with an initial position, $x$, and initial velocity, $v_x$, randomly chosen within a certain range that is larger than the acceptance. The $z$ position is initially set to zero, while $v_z$ is set to a value between $0$ and $1000$\,m/s. The bold black curve in figure~\ref{fig:acceptance} shows the acceptance expected for an ideal hexapole torus. Not surprisingly, the acceptance of a ring composed out of $1000$ straight hexapoles is virtually the same as that of an ideal ring. Note that losses due to Majorana transitions near the centre of the hexapoles are neglected, leading to an overestimate of the acceptance at low velocities (at high velocities, the atoms never cross the geometrical centre of the hexapole).

When the number of segments is decreased, two effects are observed: (i) it is seen that at high velocities the acceptance decreases, and (ii) that at specific velocities the acceptance is seen to have sharp dips. The former effect can be understood from the fact that in a straight hexapole the molecule will, on average, be further away from the geometrical centre than in a curved ring. The dips at certain velocities are so-called stop-bands~\cite{Humphries} resulting from the fact that the confinement force in a segmented ring is a periodic function. At stop-bands, the periodic force is resonant with the betatron oscillation, causing the atoms to be lost from the synchrotron as the amplitudes of their trajectories grow without bound. These resonances are absent if the length of a single hexapole is much smaller than the distance it takes for the atoms to make one transverse oscillation, i.e., if the phase-advance in a segment is $\ll 2\pi$ \footnote{Motional resonances are inevitably introduced by small misalignments of the individual hexapoles that lead to a periodic kick every round-trip.}. These stop bands were studied in a molecular synchrotron by Heiner~\emph{et al.}~\cite{Heiner2008} and Zieger~\emph{et al.}~\cite{Zieger2013a}. For the remainder of this paper we will consider a ring that consists of $40$ segments as a compromise between construction demands and the expected performance. For a $1$-meter diameter ring, this implies that each segment has a length of $78.5$\,mm. Such a ring has a radial acceptance of 2.5\,mm$\times$40\,m/s for hydrogen atoms flying at $600$\,m/s, i.e., a factor of 3 smaller than for an ideal hexapole ring with the same diameter. The vertical acceptance is 5.4\,mm$\times$100\,m/s; note however that the vertical extent of the stored beam is limited by the 1.8\,mm gap used for injecting the beam.

\subsection{Longitudinal motion in a segmented ring}
\label{sec:longitudinal}

In the electric synchrotron demonstrated by Zieger~\emph{et al.}~\cite{Zieger2010a}, bunching is achieved by temporarily ($\sim\,20\,\mu s$) increasing the voltages by $25$~$\%$ as the molecules pass the gaps between successive segments. For a magnetic ring, this would require unacceptably large currents. Therefore, we consider a scheme whereby successive segments are connected to separate power supplies which generate sinusoidally varying currents with a relative phase of $\pi$. This bunching scheme is identical to that used in linear accelerators for charged particles \cite{Humphries}. A hypothetical atom -- the so-called \textit{synchronous} atom -- flying with a velocity $v_{s}$ will pass the gaps when the currents are zero and it will keep its original velocity. Atoms that are slightly behind the synchronous atom will pass the gap when the magnetic field in the segment in front of them is reduced with respect to its nominal value, while the magnetic field in the segment behind them is increased. As a consequence, these atoms will be accelerated. Vice versa, atoms that are slightly in front of the synchronous atom will pass the gap slightly earlier when the magnetic field in the segment in front of them is larger than the magnetic field in the segment behind them and they will therefore be decelerated. As a result, atoms with a position and velocity close to those of the synchronous atom will oscillate around the synchronous atom as it revolves the ring. 

In order to calculate the longitudinal acceptance, we follow a procedure similar to the one used to describe phase stability in a Stark or Zeeman decelerator \cite{Bethlem2002,Wiederkehr2010}. When an atom moves from segment $n$ to segment $(n+1)$, it gains or loses an amount of kinetic energy that depends on the phase of the oscillating dipole field at the time that the atom passes the gap and is given by:

\begin{equation*}
\Delta K(\phi) = \mu [~(B_{\mathrm{hex,n+1}} + B_{\mathrm{dip},n+1}\sin\phi_{n+1})
\end{equation*}
\begin{equation*}
~~~~~~~~~~~~~~~~~~~~~~~~~~~~~~~~~~~~ - (B_{\mathrm{hex,n}} + B_{\mathrm{dip},n} \sin\phi_{n})~] 
\end{equation*}
\begin{equation}
= 2 \mu_{B} B_{\mathrm{dip}}\sin\phi,
\label{eq:kineticenergy}
\end{equation}

\noindent where we have used the condition that the dipole fields in successive segments have a phase difference of $\pi$, i.e., $\phi_{n+1} = \phi_{n}+\pi$, and the magnetic dipole moment of hydrogen in large fields is equal to one Bohr magneton. In order to describe the motion of non-synchronous atoms relative to the motion of the synchronous atom, we introduce the relative velocity, $\Delta v = v-v_{s}$, and phase, $\Delta \phi = \phi - \phi_{s} = \Delta z\cdot\omega/v_{s}$, with $\omega$ the frequency of the oscillating dipole field. Furthermore, the step-wise change in kinetic energy is regarded to originate from a continuously acting average force $\bar{F}(\phi)=\Delta K(\phi)/L$, where $L$ is the distance between successive bunching segments. The equation of motion is then given by

\begin{equation}
\frac{d^{2}\phi}{dt^{2}} - \frac{2\mu_{B}B_{\mathrm{dip}}}{mL} \frac{\omega}{v_{s}} (\sin\phi - \sin\phi_{s}) = 0.
\label{eq:motion}
\end{equation}

\noindent
This equation is the familiar expression for a pendulum driven by a constant torque. Atoms close to the synchronous atom will oscillate around the synchronous atom with a frequency of:

\begin{equation}
\omega_{z} = \sqrt{\frac{2\mu_{B}B_{\mathrm{dip}}}{mL}\frac{\omega}{v_s}\cos\phi_{s}}.
\label{eq:omegaz}
\end{equation}

\noindent
We will be interested only in the case of $\phi_{s}=0$, i.e., when the synchronous molecule remains at a constant speed as it revolves around the ring.

\begin{figure*}[t]
\begin{center}
\includegraphics[scale=1]{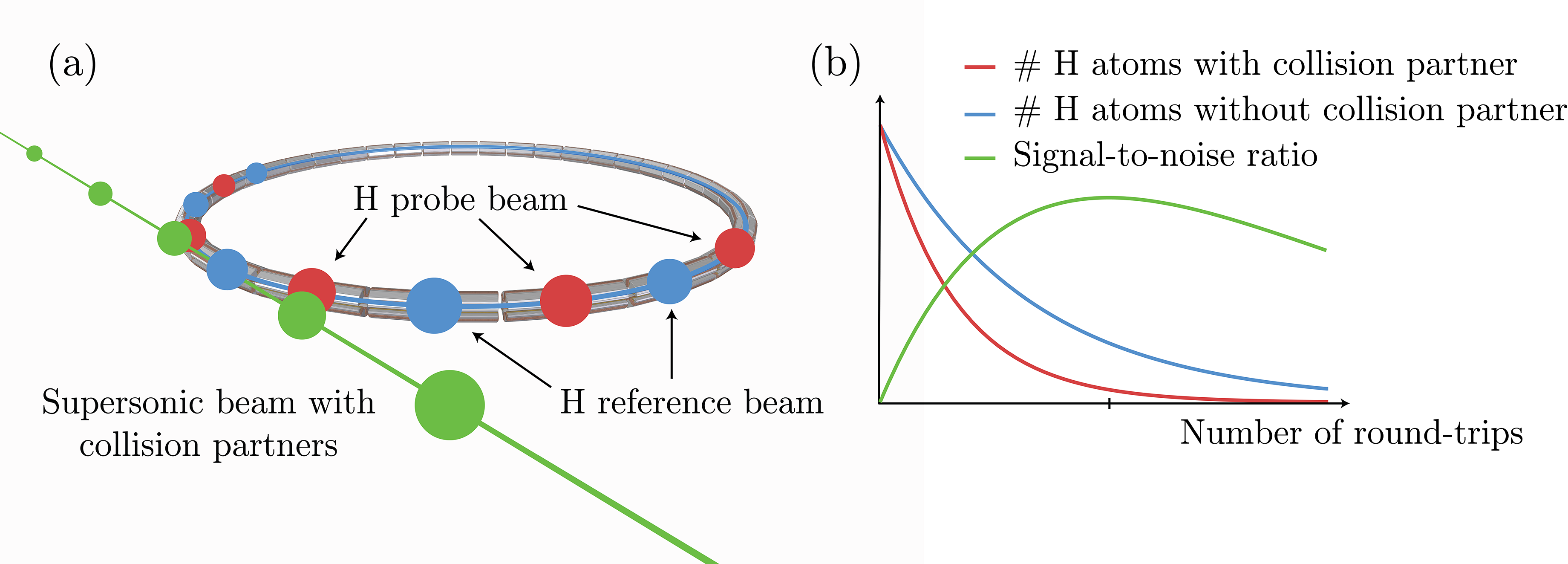}
\end{center}
\caption{(a) Schematic of the setup used for studying collisions in the synchrotron. (b) Schematic graphs of the hydrogen signal as function of the number of round-trips with (red) and without (blue) collisions, and (green) the S/N of the total (elastic + inelastic) collision cross-section that is determined from it. Here we have assumed the loss rate due to collisions $k_{col}$ to be a third of the background loss rate $k_{bg}$.}
\label{fig:collisions}
\end{figure*}

From equation~\ref{eq:omegaz}, we see that the synchrotron frequency is largest when $L$ is small, i.e., if many bunching segments are used, and if the frequency of the applied magnetic field is large. As will be discussed in section~\ref{sec:loading},  we will need to switch off at least three segments in order to load packets into the ring. Since switching between the bunching configuration shown in figure~\ref{fig:lens}(d) and the loading configuration shown in figure~\ref{fig:lens}(e) would require very sophisticated electronics, we consider the situation where every fourth segment in the ring is used for bunching, thus $L=4L_{\mathrm{hex}}$. The frequency of the oscillation, $\omega$, determines the capture range of the longitudinal well. By choosing $\omega/2\pi=5\cdot v_{s}/L_{\mathrm{hex}}=38$\,kHz the buckets will have a length of $\sim$16\,mm. Assuming an oscillating field of 5\,mT, we find a synchrotron frequency of $\omega_{z}/2\pi=42$\,Hz for hydrogen atoms with a velocity of around $600$\,m/s. The longitudinal acceptance becomes $16$\,mm by $3.0$\,m/s.

\subsection{Loading atoms in a segmented ring}
\label{sec:loading}

The use of permanent magnets makes it possible to generate strong confinement forces without the need for large currents. However, it introduces the problem of how to load the ring. The magnetic field of the hybrid hexapole lens presented in section~\ref{sec:lens} can be turned off by running a current of $380$\,A through the wires. From experiments with the electric synchrotron consisting of $40$ hexapoles, we know that switching off three segments is sufficient to allow beams to enter the synchrotron, which implies that we need a current pulse with a duration of $\sim400\,\mu$s.

Alternatively, it may also be possible to let hydrogen atoms enter the ring in a high-field-seeking state, and to use an optical transition to transfer the atoms to a low-field seeking state. For instance, a transition could be driven from the $1S_{1/2} m_{J}=-1/2$ to the $2P_{3/2} m_{J}=1/2$ state. In high field, this state decays primarily to the $1S_{1/2} m_{J}=+1/2$ state \cite{Luiten:Thesis}. Because it is not necessary to switch off the confinement fields in this scheme, it would allow loading the ring without influencing the atoms that are already stored. In this way it may be possible to reach high densities of stored hydrogen atoms. 

\begin{table*}[t]
\centering
\begin{tabular}{l = c c}
\hline
  Type of motion	    &	\multicolumn{1}{c}{Frequency}					&	Acceptance 			& Trap depth		\\
\hline
Cyclotron 			&	\Omega_{\mathrm{cycl}}/2\pi ~=~~\,190\,\mathrm{Hz}				   					\\
Synchrotron 		&	\omega_z/2\pi ~=~~~~42\,\mathrm{Hz}				&	$16$\,mm by $3.0$\,m/s 	&	$\,~0.14$\,mK  \\
Horizontal betatron &	\omega_r/2\pi ~=~2950\,\mathrm{Hz}				&	$2.5$\,mm by $40$\,m/s 	&	$24$\,mK    \\
Vertical betatron 	&	\omega_y/2\pi ~=~2950\,\mathrm{Hz}				&	$5.4$\,mm by $100$\,m/s &	$150$\,mK	\\
\hline
\end{tabular}
\label{tab:parameters}
\caption{Typical frequencies and acceptances for hydrogen atoms in the $^1S_{1/2} m_J=+1/2$ state with a forward velocity of $v_s=600$\,m/s stored in a $1$-meter diameter magnetic synchrotron consisting of $40$ hybrid lenses as shown in figure~\ref{fig:lens}. Bunching is performed by running a $5$\,A AC current with $\omega/2\pi=38$\,kHz through the electromagnets of every fourth hexapole lens. The transverse acceptances are elliptical in phase-space, the numbers given here thus correspond to their (full) axes. The longitudinal acceptance is shaped like a diamond and described by its diagonals.}
\end{table*}


\section{Studying cold collisions in a magnetic synchrotron}
\label{sec:collisions}

Our main interest for developing a magnetic synchrotron lies in its use for studying cold collisions between hydrogen atoms and various atoms and molecules\footnote{At VU University Amsterdam we are currently working on a similar experiment to study collisions between ammonia molecules stored in an electric synchrotron and molecules from a pulsed supersonic beam.}. In the envisioned experiment, schematically depicted in figure~\ref{fig:collisions}(a), $10$ packets of hydrogen atoms are made to revolve for a few seconds in the ring while colliding with pulses of atoms and molecules that are released from a high-repetition-rate pulsed valve. The hydrogen atoms are stored at a velocity of $600$\,m/s, while the velocity of the collision partner will be tuned to around $600$\,m/s by choosing the appropriate buffer gas and controlling the temperature of the pulsed valve. In this way, the collision energy can be tuned down to $0$\,K, with a resolution determined by the properties of the pulsed beam (note that, in a merged beam experiment, the energy resolution can be smaller than the energy spread of the colliding beams \cite{Shagam2013}). The packets of atomic hydrogen are injected and detected at a rate of 10\,Hz or slower, depending on the storage time. The high-repetition-rate pulsed valve running at five times the cyclotron frequency, is synchronized with the trapped packets in such a way that half of the stored packets (the probe beam) will meet with a packet from the pulsed valve during every round-trip, while the other half of the stored packets is used as a reference (the reference beam). Figure~\ref{fig:collisions}(b) shows a simulation of the expected signal in such an experiment. The blue and red curves show the number of hydrogen atoms in the reference and probe beam, respectively, as a function of the number of round-trips. Due to collisions with particles from the pulsed valve, the number of atoms in the probe beam will decay faster than that in the reference beam. From the difference in lifetime, the relative collision cross-section can be determined. In the simulation it is assumed that every (inelastic or elastic) collision leads to a loss. This is not necessarily the case at very low temperatures, when the trap depth becomes comparable to the collision energy. In order to convert the measured loss rates into a cross-section, the trap depth should be accurately measured.  

The number of collisions taking place for the atoms in the synchrotron scales with the number of round-trips, hence there will be an increase in the ratio of the number of atoms in the reference beam with respect to the number of atoms in the probe beam. At the same time, however, the number of atoms in both beams decreases, increasing the error in determining these numbers. The green line in figure~\ref{fig:collisions}(b) shows the resulting signal-to-noise ratio (S/N) for a single measurement. Initially, the S/N increases linearly with the number of round-trips until it flattens off and reaches a maximum at the number of round-trips corresponding to $\sim 2$ times the lifetime $\tau_\mathrm{probe}$ of the probe beam. For typical values of the densities and cross-sections, the lifetime of the hydrogen beams is much longer than a second, i.e., the maximum storage time for an experiment with a repetition rate of $10$\,Hz where ten packets are simultaneously stored in the ring. The repetition rate therefore needs to be reduced, lowering the number of detected atoms per second. However, the increased storage time compensates for this effectively, since the error in the measured decay rate goes as $1/\sqrt{n}\cdot 1/\tau$ for $\tau \ll 2\tau_\mathrm{probe}$, with $n$ the number of detected atoms and $\tau$ the time the atoms spend in the synchrotron. We find that, for typical densities and cross-sections, a synchrotron offers a $100$--$1000$ times enhanced S/N compared to a merged-beam experiment at a $10$\,Hz repetition rate.


\section{Conclusions}
\label{sec:conclusions}

In this paper, we have presented a design of a compact synchrotron based on $40$ straight hybrid magnetic hexapole lenses that is able to store hydrogen atoms at velocities up to $600$\,m/s. Permanent magnets are used to create strong transverse-confinement forces, while electromagnets are used for bunching and for temporarily lowering the magnetic fields to allow hydrogen atoms to enter the ring. The hexapole lenses can be mounted in a similar way as in the electric synchrotron that was demonstrated by Zieger~\emph{et al.} \cite{Zieger2010a}. The high degree of symmetry of the design ensures that the ring will be robust against misalignments and field imperfections. The electronics required to generate the necessary current pulses are routinely used in experiments with Zeeman decelerators \cite{Vanhaecke2007,Narevicius2008,Momose2013,Dulitz2014}, while cooling of the base plate thermally connected to the aluminum housing and magnet assemblies should be sufficient to take care of the modest heat load due to Ohmic losses in the coils. With analytical models that were previously used to characterize an electric synchrotron, we have calculated the oscillation frequencies and acceptance for hydrogen atoms. The results are summarised in table~\ref{tab:parameters} along with the parameters used in the simulation. The ring should be able to store a reasonable fraction of a supersonic beam at $600$\,m/s; the transverse acceptance is comparable to, while the longitudinal acceptance is about $20$ times smaller than, the phase-space occupied by atoms in a typical supersonic beam. 

A magnetic synchrotron can be used to study collisions between hydrogen (the most abundant atom in the universe) and virtually any atom or molecule, including hydrogen molecules (the most abundant molecule in the universe), He, CO and ammonia at energies below $100$\,K. This will be highly relevant for astrophysical models. As a result of the low mass of atomic hydrogen, it is expected that the (total) cross-section for collisions between hydrogen and any molecule will exhibit many resonances in this energy range, which make them exciting systems to study and will contribute to a better understanding of the processes governing collisions at low energies.


\section*{Acknowledgements}
This work is supported by the FOM-program ``Broken Mirrors \& Drifting Constants''. We thank Bas van de Meerakker, Peter Zieger, Chris Eyles and Gerard Meijer for stimulating discussions.




\end{document}